\documentclass[useAMS,usenatbib]{mn2e}
\usepackage{graphicx}

\newcommand{\kms}{\mbox{$\mathrm{km\,s^{-1}}$}}
\newcommand{\MSUN}{\mbox{$\mathrm{M_{\odot}}$}}
\newcommand{\RSUN}{\mbox{$\mathrm{R_{\odot}}$}}

\title[Accurate parameters for an ultracool white dwarf]{An accurate mass and radius measurement for an ultracool white dwarf}

\author[S. G. Parsons et al.]{S.~G.~Parsons$^{1,2}$\thanks{steven.parsons@warwick.ac.uk},
B.~T.~G{\"a}nsicke$^{1}$,
T.~R.~Marsh$^{1}$,
P~Bergeron$^{3}$,
C.~M~Copperwheat$^{1}$,
\newauthor
V.~S.~Dhillon$^{4}$,
J.~Bento$^{1}$,
S.~P.~Littlefair$^{4}$
and M.~R.~Schreiber$^{2}$
\\
$^{1}$ Department of Physics, University of Warwick, Coventry CV4 7AL, UK\\
$^{2}$ Departmento de F{\'i}sica y Astronom{\'i}a, Universidad de
Valpara{\'i}so, Avenida Gran Bretana 1111, Valpara{\'i}so, Chile\\
$^{3}$ D\'epartement de physique, Universit\'e de Montr\'eal, C.P. 6128,
Succursale Centre-Ville, Montr\'eal, QC H3C 3J7, Canada\\
$^{4}$ Department of Physics and Astronomy, University of Sheffield,
Sheffield, S3 7RH, UK}
\begin{document}
\input{references.cls}
\date{Accepted 2012 July 23.  Received 2012 July 20; in original form 
2012 June 15}

\pagerange{\pageref{firstpage}--\pageref{lastpage}} \pubyear{2012}

\maketitle

\label{firstpage}

\begin{abstract}

Studies of cool white dwarfs in the solar neighbourhood have placed a limit
on the age of the Galactic disk of 8-9 billion years. However, determining 
their cooling ages requires the knowledge of their effective temperatures,
masses, radii, and atmospheric composition. So far, these parameters could
only be inferred for a small number of ultracool white dwarfs for which an
accurate distance is known, by fitting their spectral energy distributions
(SEDs) in conjunction with a theoretical mass-radius relation. However, the
mass-radius relation remains largely untested, and the derived cooling ages
are hence model-dependent. Here we report direct measurements of the mass
and radius of an ultracool white dwarf in the double-lined eclipsing binary
SDSS\,J013851.54-001621.6. We find $M_\mathrm{WD}=0.529\pm0.010$\,\MSUN and
$R_\mathrm{WD}=0.0131\pm0.0003$\,\RSUN. Our measurements are consistent with
the mass-radius relation and we determine a robust cooling age of 9.5 
billion years for the $3570$\,K white dwarf. We find that the mass and 
radius of the low mass companion star, $M_\mathrm{sec}=0.132\pm0.003$\,\MSUN
and $R_\mathrm{sec}=0.165\pm0.001$\,\RSUN, are in agreement with evolutionary
models. We also find evidence that this $>9.5$\,Gyr old M5 star is still
active, far beyond the activity lifetime for a star of its spectral type. 
This is likely caused by the high tidally-enforced rotation rate of the 
star. The companion star is close to filling its Roche lobe and the system
will evolve into a cataclysmic variable in only 70\,Myr. Our direct 
measurements demonstrate that this system can be used to calibrate ultracool
white dwarf atmospheric models.

\end{abstract}


\section{Introduction}

White dwarfs are born hot and cool gradually over billions of years. Their 
cooling is understood well enough to make them useful in measuring the ages
of stellar populations \citep{fontaine01}. White dwarfs with brown dwarf
companions can be used to place constraints on the age of the brown dwarf
\citep{pinfield06,day-jones08}. The growing number of wide field surveys,
such as the UKIRT Infrared Deep Sky Survey (UKIDSS) and the Sloan Digital 
Sky Survey (SDSS), have lead to an increase in the number of these systems
\citep{girven10,steele11}. A proper understanding of white dwarf cooling is
essential for placing reliable limits on the ages of these systems.

White dwarf temperatures, and hence ages, are determined by fitting their
spectral energy distributions (SEDs) using models of their atmospheres. At
low temperatures ($T_\mathrm{eff}< 6000$\,K) the atmospheric models need to
include the effects of collisions between hydrogen molecules \citep{bergeron95,
bergeron02} (and with helium, if present). This effect dominates at
near-infrared wavelengths in ultracool white dwarfs ($T_\mathrm{eff}< 4000$\,K),
suppressing the infrared flux and causing it to emerge at shorter wavelengths
\citep{bergeron94}. In addition, the SEDs of ultracool white dwarfs depend on
the white dwarf's surface gravity, and hence its mass and radius. Current
white dwarf atmosphere models are not yet able to produce satisfying fits to
the observed SEDs of ultracool white dwarfs \citep{giammichele12}.

\begin{table*}
 \centering
  \caption{Journal of photometric and spectroscopic observations. Exposure times for the X-shooter observations are for the UVB, VIS and NIR arms respectively.}
  \label{obs_log}
  \vspace{2mm}
  \begin{tabular}{@{}lcccccc@{}}
  \hline
Date at     &Instrument &Filter(s) &Start  &Orbital     &Exposure     &Conditions                \\
start of run&           &          &(UT)   &phase       &time (s)     &(Transparency, seeing)    \\
\hline
2011/11/01  & ULTRACAM  & $ugr$ & 23:51 & 0.90--2.25 & 4.0         & Variable, $\sim1$ arcsec \\
2011/11/30  & RATCam    & $r$   & 20:57 & 0.81--1.07 & 10.0        & Good, $\sim1.5$ arcsec   \\
2011/11/30  & RATCam    & $i$   & 22:41 & 0.77--1.03 & 10.0        & Good, $\sim1.5$ arcsec   \\
2011/12/01  & RATCam    & $r$   & 21:24 & 0.80--1.08 & 10.0        & Good, $\sim2$ arcsec     \\
2011/12/02  & RATCam    & $r$   & 23:40 & 0.82--1.11 & 10.0        & Good, $\sim2$ arcsec     \\
2011/12/25  & X-shooter & -     & 00:58 & 0.85--2.25 & 606,294,100 & Variable, $\sim1$ arcsec \\
2012/01/08  & RATCam    & $i$   & 21:10 & 0.79--1.05 & 10.0        & Good, $\sim2$ arcsec     \\
2012/01/14  & RATCam    & $i$   & 20:34 & 0.95--1.21 & 10.0        & Good, $\sim1.5$ arcsec   \\
2012/01/18  & ULTRACAM  & $ugi$ & 19:43 & 0.29--1.47 & 4.0         & Good, $1.5-3.0$ arcsec   \\
\hline
\end{tabular}
\end{table*}

If an accurate distance is known, the absolute magnitude and a mass-radius
relation can be used in conjunction with the SED modeling to estimate
the mass and the surface gravity of the star. However, parallaxes are
available only for a small handful of ultracool white dwarfs. In one of the
best-studied cases, LHS\,3250, this method gives an unrealistically
low surface gravity of $\log g=7.27$, underlining the uncertainties in
the SED models \citep{bergeron02}. Recently \citet{kilic12} used this approach
to determine the cooling ages of the ultracool white dwarfs SDSS\,J1102+4113
and WD\,0346+246, finding cooling ages of $10{+0.4\atop-1.1}$\,Gyr and
$11.2{+0.3\atop-1.6}$\,Gyr respectively. However, they still had to rely on
the mass-radius relationship in order to determine all of their parameters.
The majority of the ultracool white dwarfs have no parallaxes, and a canonical
surface gravity of $\log g=8$ was assumed for their analysis.  However,
altering this by a plausible $\pm0.5$\,dex changes the resulting cooling age
by several Gyr \citep{kilic10b}. Furthermore, the mass-radius relation for cool
white dwarfs is all but untested observationally, further adding to the
uncertainties. Hence, at present no ultracool white dwarf has a reliable mass
determination, and hence their cooling ages are subject to large
uncertainties. 

One exception to this is the study of white dwarfs in star clusters. In
the best cases the total age of the stars is well known as is the mass at the
turnoff, hence the mass at the tip of the white dwarf cooling sequence can be
measured \citep{hansen07}. However, this picture is complicated by the presence
of binaries \citep{bedin08} and helium core white dwarfs \citep{kalirai07},
which add additional complexity to the white dwarf luminosity function.
Furthermore, these white dwarfs cannot be used to constrain the age
of the Galactic disk since they originated from a different population of
stars.

Double-lined eclipsing binaries can be used to measure masses and radii with
very few assumptions and to accuracies of better than 1 per cent \citep{
andersen91, southworth07}, independent of model atmosphere
calculations. This method has been applied to white dwarfs in eclipsing 
binaries and has resulted in the most precisely measured white dwarf
masses and radii to date \citep{parsons10, pyrzas12, parsons12}. The
subject of this paper, SDSS\,J013851.54-001621.6 (henceforth
SDSS\,0138-0016) was one of a number of candidate eclipsing white
dwarf plus main sequence binaries identified in the multi-epoch SDSS 
photometric survey, known as Stripe 82 \citep{becker11}. 

Here we present high precision photometry and spectroscopy of SDSS\,0138-0016,
confirming its binary and eclipsing nature, and use these data to directly
measure the masses and radii of both stars in the binary. The same data also
yield the temperature of the white dwarf and hence the age of the system. 

\section{Observations and their reduction}

\subsection{WHT/ULTRACAM photometry}

SDSS\,0138-0016 was observed with ULTRACAM mounted as a visitor instrument on
the $4.2$m William Herschel Telescope (WHT) on the 1st November 2011 and the
18th January 2012. ULTRACAM is a high-speed, triple-beam CCD
camera \citep{dhillon07} which can acquire simultaneous images in three
different bands; for our observations we used the SDSS $u$, $g$ and either $r$
or $i$ filters. A complete log of these observations is given in
Table~\ref{obs_log}. We windowed the CCD in order to achieve exposure times of
$\sim 4$ seconds which we varied to account for the conditions; the dead time
between exposures was $\sim 25$ ms. It is also possible to increase the
exposure time of the $u$ band observations by coadding the exposures on the
CCD before readout. Since SDSS\,0138-0016 is faint in this band ($u'=20.55$),
we used 5 coadds for the $u$ band observations, resulting in exposure times of
$\sim 20$ seconds. 

\begin{figure*}
\begin{center}
 \includegraphics[width=0.95\textwidth]{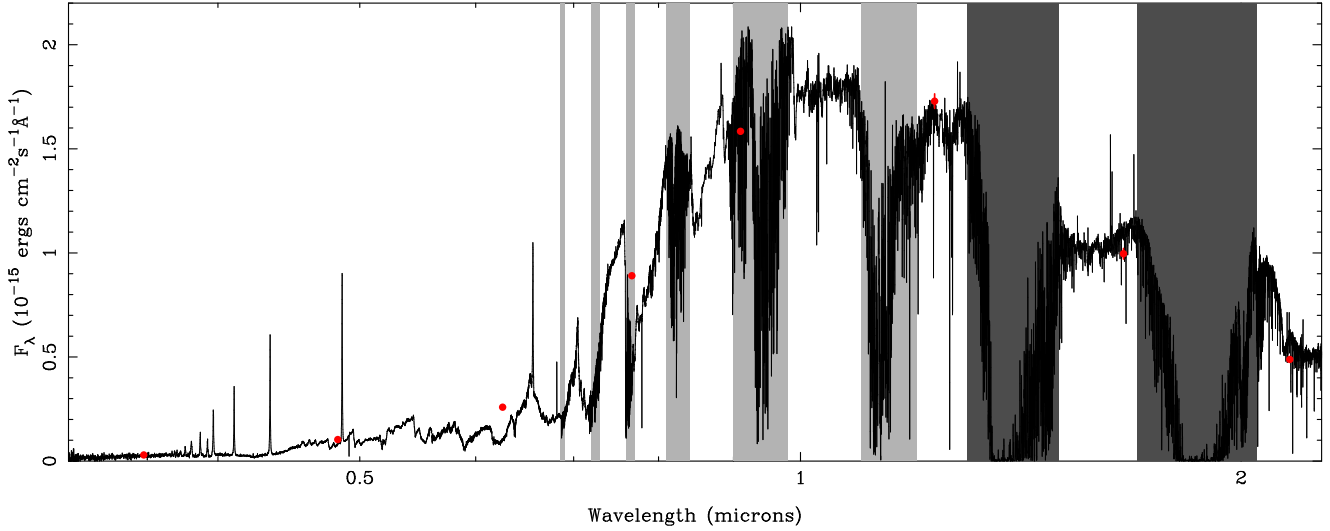}
 \caption{Averaged X-shooter spectrum of SDSS\,0138-0016, no telluric correction has been applied. Grey regions cover areas of severe (light grey) or nearly complete (dark grey) telluric absorption. The red points are the SDSS $ugriz$ and 2MASS $JHK$ magnitudes.}
 \label{av_spec}
 \end{center}
\end{figure*}

All of these data were reduced using the ULTRACAM pipeline
software. Debiassing, flatfielding and sky background subtraction were
performed in the standard way. The source flux was determined with aperture
photometry using a variable aperture, whereby the radius of the aperture is
scaled according to the full width at half maximum (FWHM). Variations in
observing conditions were accounted for by determining the flux relative to
several comparison stars in the field of view. The data were flux calibrated by
determining atmospheric extinction coefficients in each of the bands in which
we observed and we calculated the absolute flux of our target using
observations of standard stars \citep{smith02} taken in twilight. Using our
extinction coefficients we extrapolated all fluxes to an airmass of $0$. The
systematic error introduced by our flux calibration is $<0.1$ mag in all
bands. 

\subsection{LT/RATCam photometry}

Six primary eclipses of SDSS\,0138-0016 were obtained in the $r$ and $i$ bands
(three in each band) using RATCam, an optical CCD camera mounted on the
robotic $2$m Liverpool Telescope \citep{steele04}. Each eclipse observation was
composed of $75\times10$ second exposures. We used $2\times2$ binning
resulting in a readout time of $\sim5$ seconds between exposures. These
observations are summarised in Table~\ref{obs_log}.

The raw data are automatically run through a pipeline that debiases, removes a
scaled dark frame and flat-fields the data. The source flux was determined
with aperture photometry using the ULTRACAM pipeline. The same nearby stars
used to flux calibrate the ULTRACAM data were used to calibrate the RATCam
data. 

\subsection{VLT/X-shooter spectroscopy}

We obtained service mode observations of SDSS\,0138-0016 with
X-shooter \citep{dodorico06} mounted at the VLT-UT2 telescope. The observations
were designed to cover an entire orbit of the system. Details of these
observations are listed in Table~\ref{obs_log}. X-shooter is a medium resolution
spectrograph consisting of 3 independent arms that give simultaneous spectra
longward of the atmospheric cutoff (0.3 microns) in the UV (the ``UVB'' arm),
optical (the ``VIS'' arm) and up to 2.5 microns in the near-infrared (the
``NIR''arm). We used slit widths of 1.0'', 0.9'' and 0.9'' in X-shooter's
three arms and binned by a factor of two in the dispersion direction in the
UVB and VIS arms resulting in a spectral resolution of 2500--3500 across the
entire spectral range.

The reduction of the raw frames was conducted using the standard pipeline
release of the X-shooter Common Pipeline Library (CPL) recipes (version 1.3.7)
within ESORex, the ESO Recipe Execution Tool, version 3.9.0. The standard
recipes were used to optimally extract and wavelength calibrate each
spectrum. The instrumental response was removed by observing the
spectrophotometric standard star Feige\,110 and dividing it by a flux table of
the same star to produce the response function. The wavelength scale was also
heliocentrically corrected.  

\section{Results}

\begin{table}
 \centering
  \caption{Identified white dwarf emission lines. $\gamma_\mathrm{WD}$
  is the systemic velocity of the line and $K_\mathrm{WD}$ is the measured
  radial velocity of the line.}
  \label{emis_lines}
  \begin{tabular}{@{}lccc@{}}
  \hline
Line         & Wavelength & $\gamma_\mathrm{WD}$ & $K_\mathrm{WD}$  \\
             & (\AA)      & (\kms)              & (\kms)          \\ 
\hline
H\,14        & $3721.948$ & $108.5\pm3.9$       & $82.1\pm4.2$    \\
H\,13        & $3734.372$ & $115.6\pm7.3$       & $74.6\pm9.5$    \\
Fe\,{\sc i}  & $3737.131$ & $106.5\pm3.3$       & $87.0\pm4.6$    \\
Fe\,{\sc i}  & $3745.899$ & $108.6\pm2.8$       & $85.3\pm4.1$    \\
H\,12        & $3750.152$ & $109.1\pm5.3$       & $83.6\pm7.7$    \\
H\,11        & $3770.634$ & $104.4\pm3.8$       & $84.4\pm5.2$    \\
H\,10        & $3797.910$ & $105.7\pm6.0$       & $90.4\pm8.8$    \\
H\,9         & $3835.397$ & $111.7\pm7.6$       & $78.8\pm9.0$    \\
H\,8         & $3889.055$ & $100.4\pm1.5$       & $81.5\pm2.5$    \\
Ca\,{\sc ii} & $3933.663$ & $105.0\pm1.0$       & $86.5\pm1.0$    \\
Ca\,{\sc ii} & $3968.469$ & $106.6\pm1.8$       & $85.0\pm2.8$    \\
H$\epsilon$  & $3970.074$ & $103.8\pm1.1$       & $87.0\pm1.3$    \\
H$\delta$    & $4101.735$ & $106.8\pm1.1$       & $87.5\pm1.3$    \\
Ca\,{\sc i}  & $4226.728$ & $101.8\pm8.3$       & $80.4\pm4.9$    \\
H$\gamma$    & $4340.465$ & $106.4\pm1.0$       & $87.3\pm1.0$    \\
Fe\,{\sc i}  & $4383.545$ & $104.6\pm1.8$       & $84.0\pm2.6$    \\
H$\beta$     & $4861.327$ & $106.5\pm0.5$       & $87.0\pm0.6$    \\
Mg\,{\sc i}  & $5167.322$ & $110.6\pm6.6$       & $88.8\pm7.9$    \\
Mg\,{\sc i}  & $5172.684$ & $105.1\pm1.7$       & $87.1\pm2.6$    \\
Mg\,{\sc i}  & $5183.604$ & $102.0\pm1.2$       & $85.9\pm1.8$    \\
H$\alpha$    & $6562.760$ & $104.5\pm0.3$       & $85.5\pm0.4$    \\
\hline
\end{tabular}
\end{table}

\subsection{Radial Velocities}

\begin{figure}
\begin{center}
 \includegraphics[width=0.95\columnwidth]{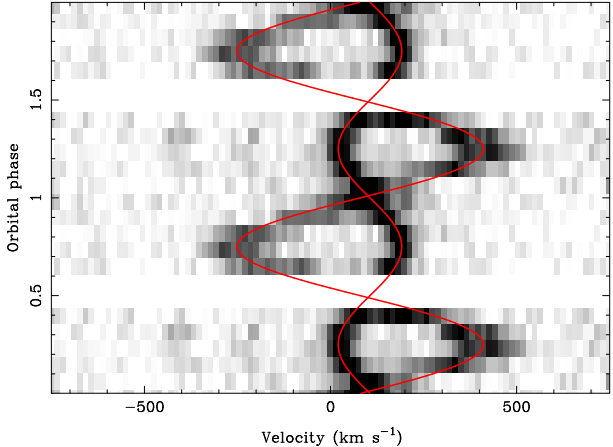}
 \caption{A trailed spectrum of the Ca\,II 3934\,{\AA} line. Emission
can be seen from both components. The strongest component is from the white
dwarf's chromosphere whilst the weaker (but larger amplitude) component
originates from the main-sequence star. The red lines (online version only)
track the motion of each component. Due to the short duration of the eclipse,
the spectra taken at phase zero still covered the out-of-eclipse phase hence
we are unable to say whether the white dwarf emission component is eclipsed.
There was a 10 minute gap in observations around phase 0.5, which was made
for calibration reasons.}
 \label{ca_trail}
 \end{center}
\end{figure}

Figure~\ref{av_spec} shows the average spectrum for SDSS\,0138-0016. The M 
star features dominate the spectrum, but there are also several emission lines
that move in anti-phase to the absorption features of the M star. These 
emission lines originate from the white dwarf's chromosphere as a result of
accretion of material from the wind of the M star. They have been seen in 
other close white dwarf plus main sequence binaries and reliably track the 
motion of the white dwarf \citep{tappert11b,tappert11a}. A list of the 
unambiguously detected emission lines from the white dwarf is given in 
Table~\ref{emis_lines}, though there are likely to be additional lines at 
longer wavelengths which are obscured by the dominant M star. A trailed 
spectrum of the Ca\,{\sc ii} 3934\,{\AA} line is shown in
Figure~\ref{ca_trail}.

\begin{figure}
\begin{center}
 \includegraphics[width=0.95\columnwidth]{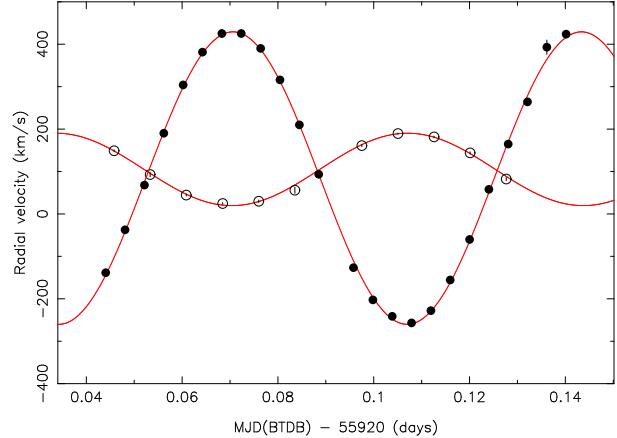}
 \caption{Radial velocity fits to the Ca\,{\sc ii} 3934\,{\AA} emission line from the white dwarf (open circles) and the K\,{\sc i} 7700\,{\AA} absorption line from the M star (filled circles).}
 \label{rvs}
 \end{center}
\end{figure}

Each emission line was fitted with a combination of a first order polynomial
and a Gaussian component. For all the Balmer lines and the Ca\,{\sc ii} lines
there is also an emission component from the M star due to activity, in all
cases it is the emission component from the white dwarf that is stronger 
(see Figure~\ref{ca_trail} for example). For these lines we fit both 
components simultaneously using a combination of a polynomial and two 
Gaussians. For the M star, we fit the K\,{\sc i} 7700\,{\AA} absorption line.
Although fits to other M star absorption features give consistent results, 
the K\,{\sc i} 7700\,{\AA} line is the cleanest feature available. 
Figure~\ref{rvs} shows sinusoidal fits to the measured radial velocities for 
both the white dwarf and the M star. Table~\ref{emis_lines} lists the fitted
radial velocities for the white dwarf emission lines. We measure 
$K_1=86.5\pm1.0$\,\kms and $K_2=346.7\pm2.3$\,\kms. The two radial velocity 
amplitudes give the mass ratio of the two stars 
$q=M_\mathrm{sec}/M_\mathrm{WD}=0.249\pm0.003$. 

\subsection{Light curve model fitting}

\begin{figure*}
\begin{center}
 \includegraphics[width=0.97\textwidth]{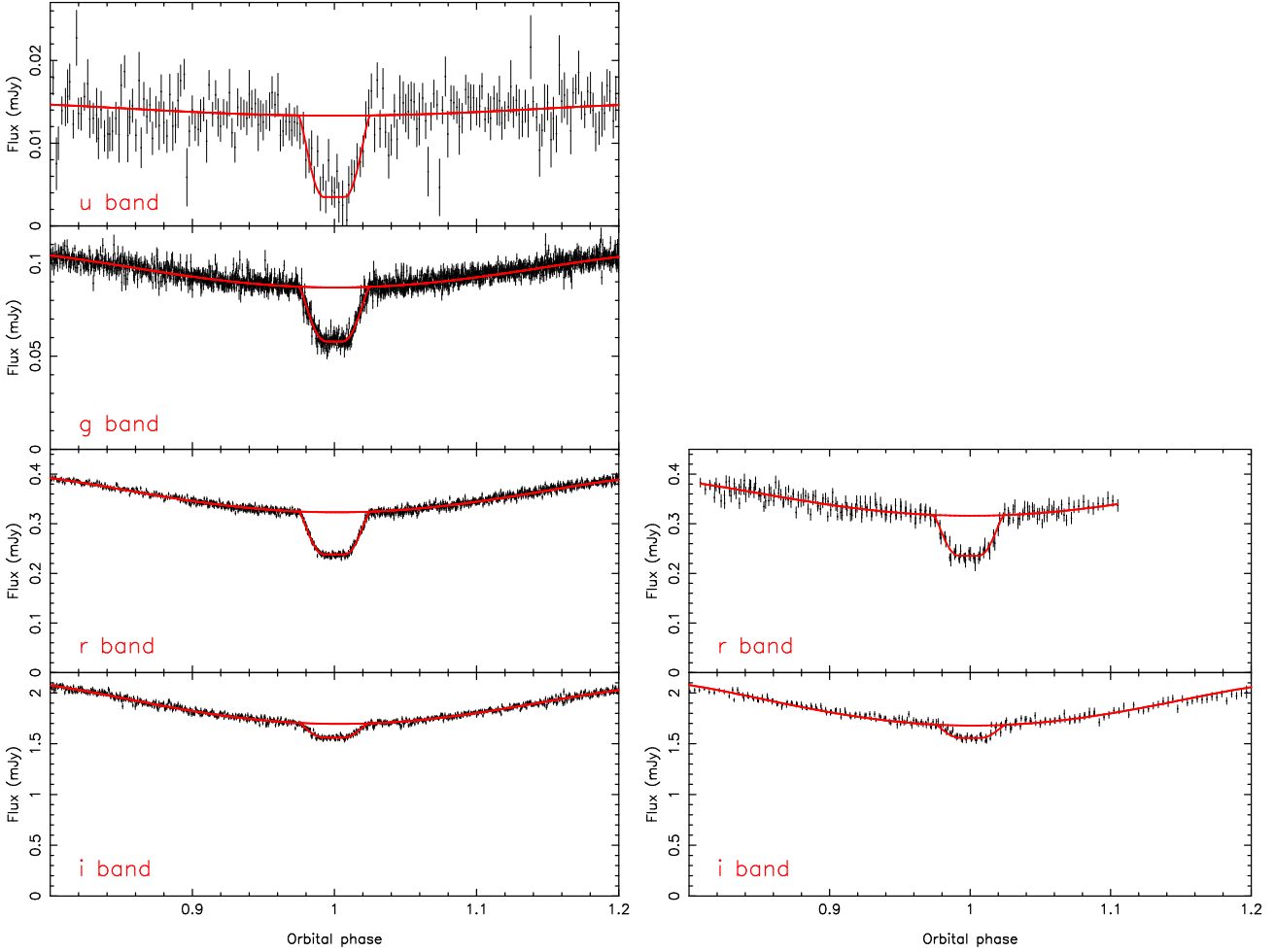}
 \caption{Primary eclipse light curves and fits. The data shown in the
left hand panels were obtained using ULTRACAM, whilst the right hand panels
are from RATCam. Two models are shown for each light curve, the best fit model
and the same model but with no primary eclipse, demonstrating the relative
contribution of each star in the different bands.}
 \label{lc_fit}
 \end{center}
\end{figure*}

Figure~\ref{lc_fit} shows our light curves of SDSS\,0138-0016 around the 
expected time of the eclipse of the white dwarf by the M star. Our data
confirm the eclipsing nature of the system. The reduced depth of the eclipse
at longer wavelengths confirms that the bluer white dwarf is being eclipsed.

To measure the system parameters we fitted all the light curves using a code
written for the general case of binaries containing white
dwarfs \citep{copperwheat10}. It has been used in the study of other white
dwarf-main sequence binaries \citep{parsons10, pyrzas12, parsons12}. The
program subdivides each star into small elements with a geometry fixed by its
radius as measured along the direction of centres towards the other star,
Roche geometry distortion and beaming are also included. The code also
calculates the white dwarf contribution to the overall flux. 

The parameters needed to define the model were: the mass ratio, $q =
M_\mathrm{sec}/M_\mathrm{WD}$, the inclination, $i$, the stellar radii scaled
by the orbital separation $R_\mathrm{sec}/a$ and $R_\mathrm{WD}/a$, quadratic
limb darkening coefficients for both the stars, the time of mid eclipse,
$T_{0}$, the period, $P$ and flux scaling factors for each star. 

The primary eclipse shape does not contain enough information to determine the
inclination and scaled radii of both stars simultaneously. However, the
amplitude of the ellipsoidal modulation is related to the mass ratio and
$R_\mathrm{sec}/a$. Therefore, since we knew the mass ratio from our
spectroscopic observations we could use it as a prior constraint and hence we
were able to measure these parameters simultaneously. Unfortunately, this
approach did not work for the $u$ band data since the white dwarf dominates
the overall flux in this band, suppressing the ellipsoidal modulation and 
making it much harder to fit. It is also of much lower signal-to-noise and
may be affected by activity from the M star. Since the RATCam light curves
only covered the primary eclipse they could not be used to determine accurate 
parameters. However, we used them to measure the magnitudes of the two stars.
We fitted all the RATCam light curves separately using the parameters found
from the ULTRACAM light curves and allowed the flux scaling factors to vary.

For fitting the light curves, we phase folded the data and kept the period
fixed as unity. The limb darkening of both stars was set using a 4-coefficient
formula: 

\begin{equation}
I(\mu)/I(1) = 1-\sum\limits_{i=1}^4 a_i(1-\mu^{i/2}),
\end{equation}
where $\mu$ is the cosine of the angle between the line of sight and the
surface normal. For the secondary star, we use the coefficients for a
$T_\mathrm{eff}=2900,\, \log{g}=5$ main sequence star \citep{claret11}. For the
white dwarf we calculated the 4 coefficients from a white dwarf model
atmosphere with $T_\mathrm{WD}=3500$ and $\log{g}=7.9$, folded through the
appropriate filter profiles. We kept all limb darkening parameters fixed. 

\begin{table}
 \centering
  \caption{Light curve model parameters from Markov chain Monte Carlo minimisation. The limb darkening coefficients $(a_i)$, are also listed for each star.}
  \vspace{2mm}
  \label{lc_params}
  \begin{tabular}{@{}lccc@{}}
  \hline
 Parameter & $g$ & $r$ & $i$ \\
 \hline 
$i$                  & $77.14^\circ\pm0.04^\circ$ & $77.19^\circ\pm0.02^\circ$ & $77.25^\circ\pm0.07^\circ$ \\
$r_\mathrm{WD}/a$     & $0.0200\pm0.0007$         & $0.0205\pm0.0005$         & $0.0212\pm0.0018$        \\
$r_\mathrm{sec}/a$    & $0.333\pm0.001$           & $0.330\pm0.001$           & $0.328\pm0.001$          \\
$a_1\,(\mathrm{WD})$ & $-0.129$          & $1.185$           & $1.499$            \\
$a_2\,(\mathrm{WD})$ & $2.710$           & $-0.473$          & $-0.402$           \\
$a_3\,(\mathrm{WD})$ & $-3.079$          & $-0.041$          & $-0.457$           \\
$a_4\,(\mathrm{WD})$ & $1.161$           & $0.096$           & $0.305$            \\
$a_1\,(\mathrm{sec})$& $0.2083$          & $0.5288$          & $0.6659$           \\
$a_2\,(\mathrm{sec})$& $1.1341$          & $0.2451$          & $0.5135$           \\
$a_3\,(\mathrm{sec})$& $-0.4029$         & $0.4560$          & $-0.2498$          \\
$a_4\,(\mathrm{sec})$& $0.0467$          & $-0.2531$         & $0.0307$           \\
 \hline
\end{tabular}
\end{table} 

We used the Markov Chain Monte Carlo (MCMC) method to determine the
distributions of our model parameters \citep{press07}. The MCMC method involves
making random jumps in the model parameters, with new models being accepted or
rejected according to their probability computed as a Bayesian posterior
probability. In this instance this probability is driven by a combination of
$\chi^2$ and the prior probability from our mass ratio constraint. 
Table~\ref{lc_params} lists the best fit parameters and their statistical
errors, along with the limb darkening coefficients used for both stars. The
fits to the $g$, $r$ and $i$ band light curves all give consistent
results. Figure~\ref{lc_fit} shows the fits to each band around the primary
eclipse. We also show the best fitting models but with the primary eclipse
turned off, to illustrate the contribution of each star in the various bands.

\begin{figure}
\begin{center}
 \includegraphics[width=0.98\columnwidth]{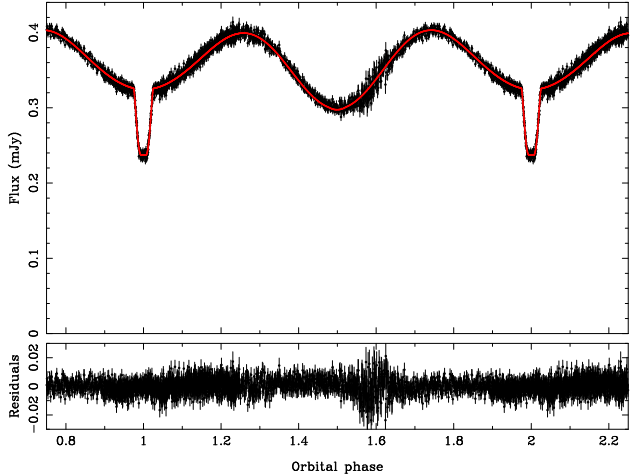}
 \caption{ULTRACAM $r$ band full orbit light curve of SDSS\,0138-0016.
The red line (online version only) shows the best fit to the data. The 
out-of-eclipse variations are caused by the tidally distorted main-sequence
star presenting a different surface area during the orbit. The amplitude of
this modulation is related to the mass ratio ($q=M_\mathrm{sec}/M_\mathrm{WD}$) 
and the radius of the main-sequence star. Since we know the mass ratio from
the spectroscopy, the amplitude of the modulation, combined with the eclipse
shape, allows us to measure the orbital inclination and radii of both stars. 
The lower panel shows the residuals to the fit, a small region of data was
affected by clouds (around phase 1.6).}
 \label{full_fit}
 \end{center}
\end{figure}

The best-fit model to the full light curve is shown in Figure~\ref{full_fit}.
We find an inclination of $77.19^\circ\pm0.02^\circ$ and a white dwarf mass
and radius of $0.529\pm0.010$\,{\MSUN} and $0.0131\pm0.0003$\,{\RSUN} 
respectively. The surface gravity of the white dwarf is then 
log $g=7.926\pm0.022$. 

\section{White dwarf temperature and age}

\begin{figure}
\begin{center}
 \includegraphics[height=\columnwidth,angle=270]{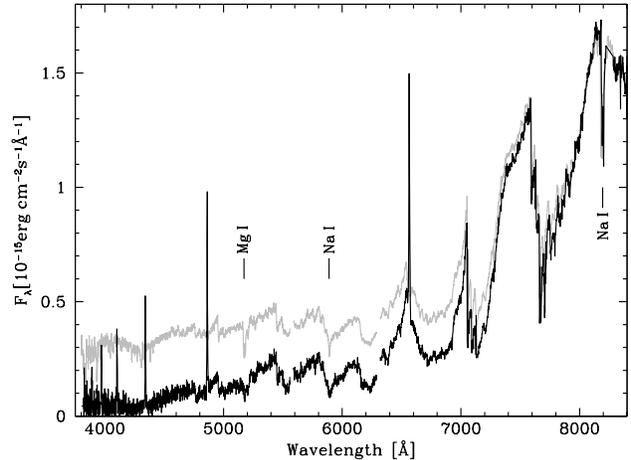}
 \caption{SDSS spectra of SDSS\,0138-0016 (black line) and the similar eclipsing white dwarf plus main-sequence binary SDSS\,1210+3347 (gray line) which contains a $6000$\,K white dwarf. Both of these systems contain an M5 main sequence star and their spectra have been scaled to match in the $i$ band. Therefore, any discrepancy at shorter wavelengths reflects differences in the white dwarf components.}
 \label{compare}
 \end{center}
\end{figure}

Figure~\ref{compare} shows the SDSS spectrum of SDSS\,0138-0016 in black and
that of a second eclipsing white dwarf plus main-sequence binary 
SDSS\,1210+3347 in gray \citep{pyrzas12}. These two systems are very similar
in that they contain M5 main-sequence stars and cool white dwarfs. Despite
the low temperature of $6000\pm200$\,K for the white dwarf in SDSS\,1210+3347
a substantial blue excess is still produced \citep{pyrzas12}. No such excess
is seen in the spectrum of SDSS\,0138-0016 implying that the white dwarf is 
much cooler than the one in SDSS\,1210+3347.

From our light curve fits we measure white dwarf magnitudes of 
$g=20.242\pm0.007$, $r=19.079\pm0.006$ and $i=18.773\pm0.020$. We also measure
$u=21.42\pm0.06$, however, the $u$ band magnitude is likely to be unreliable
since we know that the M star is active and any activity on the M star will
heavily affect the $u$ band due to Balmer continuum emission, therefore we do
not use the $u$ band magnitude to constrain the temperature of the white dwarf.

\begin{figure}
\begin{center}
 \includegraphics[width=0.98\columnwidth]{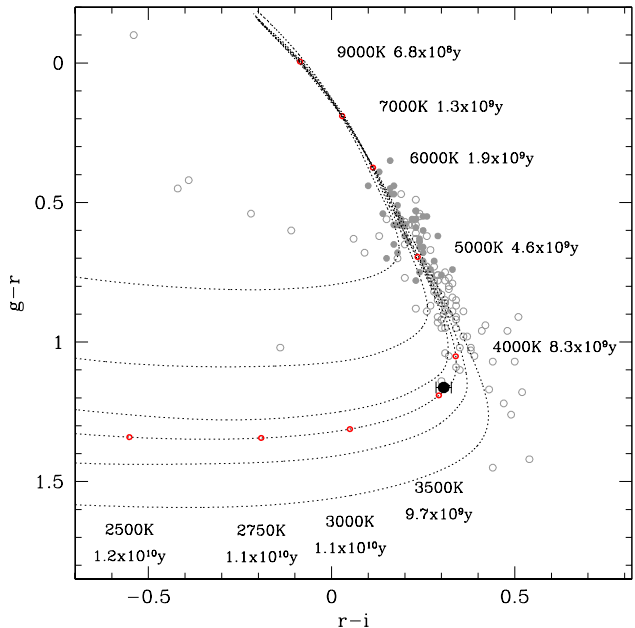}
 \caption{Colour-colour plot of cool white dwarfs. The dotted lines are
cooling tracks for different atmospheric compositions, from top to bottom
$\log(\mathrm{He/H})=2.0, 1.0, 0.5, 0.3, 0.0, -1.0$. In the calculation of
these tracks, the mass of the white dwarf was fixed to 0.529\,{\MSUN}, as
determined from our spectroscopic and photometric fits. The best model
atmosphere fit is found for $\log(\mathrm{He/H})=0.3$ and
$T=3570{+110\atop-80}$\,K. Cooling ages and temperatures are given for the
$\log(\mathrm{He/H})=0.3$, and we determine the age of the white dwarf in
SDSS\,0138-0106 to be $9.5{+0.2\atop-0.3}$\,Gyr. Furthermore, the surface
gravity for this solution is in agreement with our measured value, meaning
that the radius of the white dwarf in SDSS\,0138-0016  is completely
consistent with current mass-radius relationships.}
 \label{wd_cols}
 \end{center}
\end{figure}

The colours of the white dwarf in SDSS\,0138-0016 are shown in
Figure~\ref{wd_cols} along with those of other cool and ultracool white
dwarfs \citep{harris99, harris01, hall08, kilic10a, kilic10b, leggett11}.
We computed a set of model atmospheres spanning a wide
range in effective temperatures and atmospheric He/H abundance
ratios \citep{giammichele12}, we also include the opacity from the red wing of
$L\alpha$. In all cases the mass was kept fixed as $0.529$\,\MSUN. For low
temperatures ($<4000$\,K), the evolution of the cooling tracks in the $r-i$
colour depends almost exclusively on the temperature, and in $g-r$ on the He/H
abundance ratio. Ultracool white dwarfs become bluer in $r-i$ with decreasing
temperature, because of collisionally induced absorption, and bluer in $g-r$
with increasing He/H ratio. The colours of the white dwarf in
SDSS\,0138-0016 therefore unambiguously constrain both its
temperature, $T=3570{+110\atop-80}$\,K, and its atmospheric
composition, $\log(\mathrm{He/H})=0.3$. The He abundance seems
plausible, as we know that the white dwarf is accreting from the wind
of its H-rich companion, and that the accreted material will be
mixed within the deep convection zone that is typical for very cool
white dwarfs \citep{dufour07}. The model that best fits the measured colours
of the white dwarf gives a surface gravity consistent with the value obtained
from our light curve fits. This internal consistency strongly supports the
validity of our results and also independently confirms the accuracy of the
mass-radius relationship of cool white dwarfs. Our final best-fit model to
the temperature and helium abundance implies a cooling age of 
$9.5{+0.2\atop-0.3}$\,Gyr for the white dwarf in SDSS\,0138-0016, 
making it one of the oldest white dwarfs with an accurate cooling age.

\section{Distance and kinematics}

We followed the prescription of \citet{beuermann06} to estimate the distance of
SDSS\,0138-0016. For M-dwarfs, the surface brightness near 7500\,\AA, and
depth of the TiO band near 7165\,\AA\ are a strong function of the spectral
type. \citet{beuermann06} provides a calibration of the surface brightness
$F_\mathrm{TiO}$ defined as the the difference between the mean surface fluxes
in the bands 7450-7550\,\AA\ and 7140-7190\,\AA. Measuring the observed flux
$f_\mathrm{TiO}$ from the spectrum, the distance is then calculated as

\begin{equation}
d=\sqrt{R_\mathrm{sec}^2\frac{F_\mathrm{TiO}}{f_\mathrm{TiO}}}.
\end{equation}

Given the high accuracy of $R_\mathrm{sec}$ determined from the light curve
model, the main uncertainties in the distance estimate are the flux
calibration of the spectroscopy, and the spectral type of the
companion. Adopting a conservative uncertainty in the spectral type of
M$5\pm1.0$, and using the SDSS and the VLT/X-shooter spectrum, we find a
distance of $52{+13\atop-10}$\,pc, with the uncertainty in the spectral type
dominating the error balance. Since the shape of the M-dwarf in SDSS\,0138-0016
is distorted by the white dwarf (see Section~\ref{sec:params}) we calculate 
the distance using the measured radius in several directions (and the spectra
taken at the corresponding phase at which that radius is visible). We find
that the oblateness of the M dwarf has a minor affect on the measured distance,
much smaller than the uncertainty in the spectral type. The absolute magnitude
of the white dwarf implied by this distance is $M_g=16.66{+0.46\atop-0.48}$, 
which is in agreement with the best-fit SED model, $M_g=16.89$, further 
validating our results. 

SDSS\,0138-0016 has a relatively large proper motion 
($\mu_{\alpha}=336.2\pm3.8$\,mas$/$yr, $\mu_{\delta}=32.9\pm3.8$\,mas$/$yr)
and a systemic velocity, measured from the X-shooter spectra, of 
$84.5\pm1.2$\,\kms. Using these values and the measured distance gives 
space velocities of  $U=-99\pm9$\,\kms, $V=+198\pm6$\,{\kms} and
$W=-45\pm3$\,\kms, which makes it likely that SDSS\,0138-0016 is a 
member of the thick disk \citep{pauli06}. 

\section{Discussion} 

\subsection{System parameters} \label{sec:params}

A full list of the physical parameters of SDSS\,0138-0016 is given in 
Table~\ref{sys_paras}. The secondary star's shape is highly distorted due to 
the presence of the nearby white dwarf. It fills $91\%$ of its Roche lobe (as
measured towards the L1 point). Therefore Table~\ref{sys_paras} lists the 
radius of the secondary star in various directions. For our final discussions 
we adopt the volume-averaged radii.

\begin{table}
\caption{System parameter of SDSS\,0138-0016.}
\label{sys_paras}
\begin{tabular}{@{}ll@{}}
\hline
Parameter                         & Value                    \\
\hline
Orbital period                    & $0.072\,764\,91(2)$\,days\\
T0 (MJD(BTDB))                    & $55867.007405(6)$        \\
Orbital separation                & $0.639\pm0.004$\,\RSUN   \\
Orbital inclination               & $77.19^\circ\pm0.02^\circ$  \\
White dwarf mass                  & $0.529\pm0.010$\,\MSUN   \\
White dwarf radius                & $0.0131\pm0.0003$\,\RSUN \\
White dwarf log $g$               & $7.926\pm0.022$          \\
White dwarf effective temperature & $3570{+110\atop-80}$\,K  \\
White dwarf cooling age           & $9.5{+0.2\atop-0.3}$\,Gyr\\
Secondary star mass               & $0.132\pm0.003$\,\MSUN   \\
Secondary star radius sub-stellar & $0.211\pm0.001$\,\RSUN   \\
Secondary star radius polar       & $0.157\pm0.001$\,\RSUN   \\
Secondary star radius backside    & $0.183\pm0.001$\,\RSUN   \\
Secondary star radius side        & $0.163\pm0.001$\,\RSUN   \\
Secondary star radius volume-averaged & $0.165\pm0.001$\,\RSUN   \\
Distance                          & $52{+13\atop-10}$\,pc    \\
\hline                                                                  
\end{tabular}
\end{table}

\begin{figure}
\begin{center}
 \includegraphics[width=0.98\columnwidth]{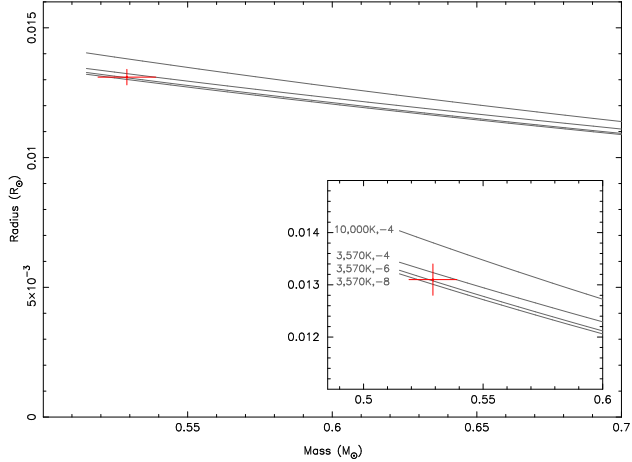}
 \caption{Mass-radius plot for the white dwarf in SDSS\,0138-0016. The gray lines are theoretical mass-radius tracks from \citet{benvenuto99} where the first number is the temperature of the white dwarf and the second number is the exponent of the hydrogen layer fraction.}
 \label{wdmr}
 \end{center}
\end{figure}

Figure~\ref{wdmr} show the mass-radius plot for the white dwarf in SDSS\,0138-0016. Also plotted are a number of theoretical mass-radius tracks from \citet{benvenuto99}. Our measurements are in excellent agreement with the models, although not precise enough to distinguish between the different hydrogen layer masses. Nevertheless, this consistency reinforces our temperature and age measurement. 

\begin{figure}
\begin{center}
 \includegraphics[width=0.98\columnwidth]{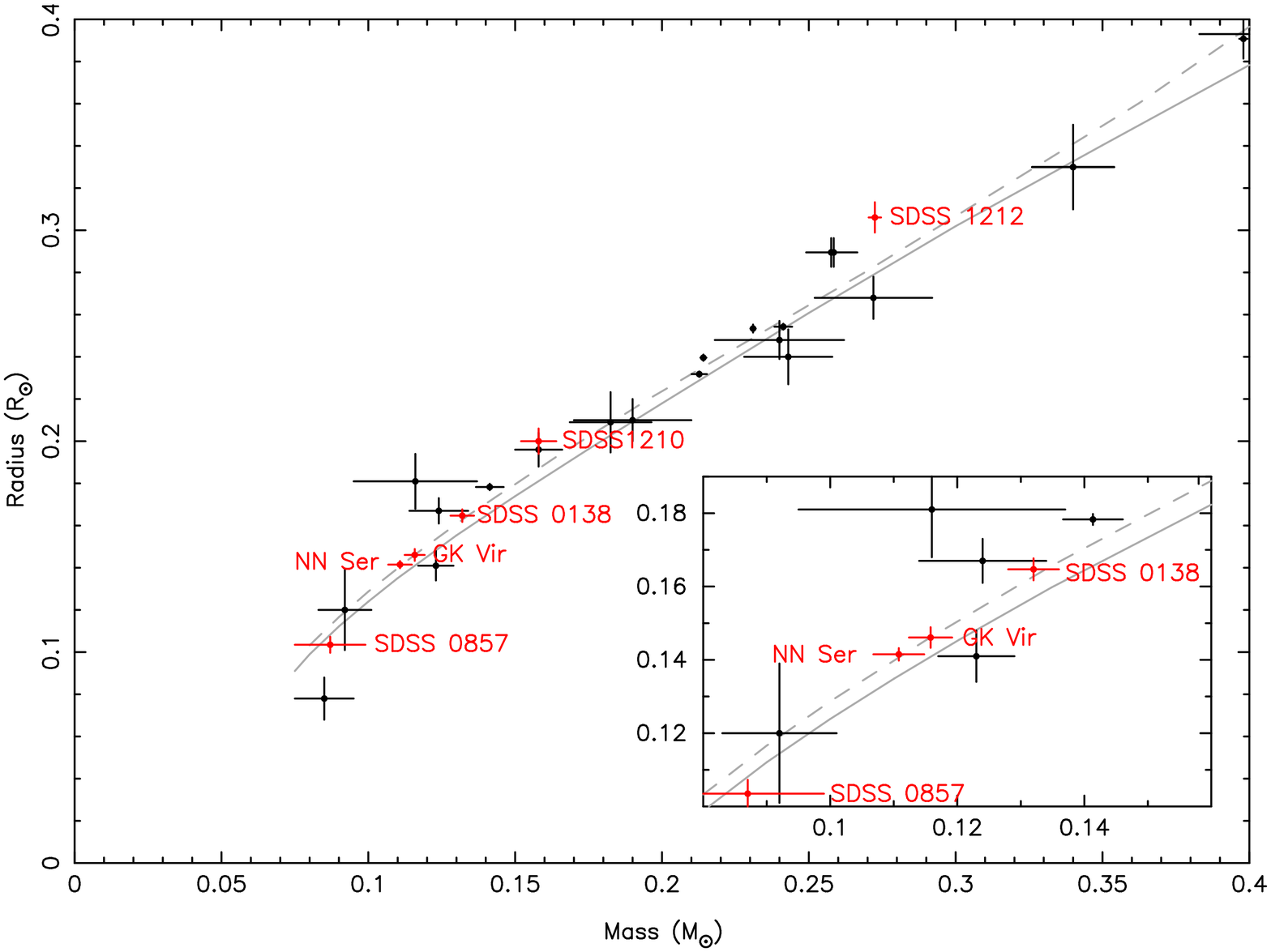}
 \caption{Mass-radius plot for low mass stars. The mass and radius values for the M star in SDSS\,0138-0016 are shown as well as others from eclipsing white dwarf binaries in red \citep{parsons10, pyrzas12, parsons12,parsons12b}. Other measurements are from \citet{knigge11, carter11, ofir12}. The solid line is the 8\,Gyr isochrone from \citet{baraffe98} whilst the dashed line is a 5\,Gyr model from \citet{morales10} which includes the effects of magnetic activity. Also shown is a zoom in on the region around SDSS\,0138-0016.}
 \label{mdmr}
 \end{center}
\end{figure}

The mass-radius plot for low-mass stars is shown in Figure~\ref{mdmr}. The measured mass and radius of the low-mass star in SDSS\,0138-0016 are consistent with evolutionary models. Also shown are a number of other precise mass-radius measurements from low-mass stars that are in eclipsing binaries with white dwarfs. The precision of these measurements demonstrates the potential of these systems for testing low-mass stellar models. The number of these systems has increased rapidly in the last few years \citep{steinfadt08,pyrzas09,pyrzas12,nebot09,drake09,drake10,law11} making them a valuable resource for testing the mass-radius relationship for low-mass stars.

As previously noted, the X-shooter spectra of SDSS\,0138-0016 show emission components in the Balmer and Ca\,{\sc ii} lines originating from the M star due to activity (in addition to the components from the white dwarf). This indicates that the M star is still active, despite its age. The M5 star has an age of least 9.5\,Gyr, likely more than 10\,Gyr when the main-sequence lifetime of the white dwarf progenitor is taken into account. \citet{west08} list the activity lifetime of an M5 star as $7.0\pm0.5$\,Gyr, substantially shorter than the age of the M star in SDSS\,0138-0016. Therefore, it is likely that the tidally-induced rapid rotation of the M star keeps it active and makes it appear younger.

\subsection{Evolution of the system}

We reconstruct the past and future evolution of SDSS\,0138-0016 using the
tools described in \citet{schreiber03} and \citet{zorotovic10, zorotovic11}.

Assuming that the only mechanism of angular momentum loss from the system is
via gravitational radiation then SDSS\,0138-0016 emerged from the common
envelope 9.5\,Gyr ago with an orbital period of 5.28 hours.

Fixing the common envelope efficiency to $\alpha=0.25$ results in a mass of
the white dwarf progenitor of 1.83\MSUN. The evolutionary time scale of the
white dwarf progenitor in this case would have been 1.63\,Gyr, giving a total
age of the system of 11.13\,Gyr. Allowing for values of $\alpha$ between 0 and
1 but insisting that the evolutionary timescale of the progenitor is less than
4\,Gyr (i.e. the system must be younger than 13.5\,Gyr), leads to a range of
progenitor masses between 1.39 and 2.00\MSUN.

A total age of $\sim$$11$\,Gyr makes it more likely that SDSS\,0138-0016 is a 
member of the thick disk (as the space motion implies). This makes it a 
fairly old member of this population but consistent with previous kinematic 
studies that found that thick disk stars have a mean age of $10\pm2$\,Gyr
\citep{feltzing09}. However, there is still some uncertainty as to whether
the Galaxy has a thick-thin disk bi-modality \citep{bovy12}.

The continuing loss of orbital angular momentum will lead SDSS\,0138-0016
to become a cataclysmic variable in 70\,Myr at which point it will have an
orbital period of 1.66 hours. Due to the long angular momentum loss time-scale,
systems of this type, so close to mass transfer, are likely to be rare. 

\section{Conclusions}

Using high-precision photometric and spectroscopic data we measure the mass and radius of the ultracool white dwarf and low-mass star in the eclipsing binary SDSS\,0138-0016. We use this information and the colour of the white dwarf to determine its atmospheric composition, temperature and age. We find that the white dwarf has a temperature of $3570{+110\atop-80}$\,K and has been cooling for $9.5{+0.2\atop-0.3}$\,Gyr. We also find that the mass and radius measurements for both the ultracool white dwarf and the low-mass star are consistent with evolutionary models. This supports the use of theoretical white dwarf mass-radius relationships when attempting to determine the temperature of ultracool white dwarfs using SED fitting and parallax measurements.

We find that the activity lifetime of the main-sequence star has been greatly extended due to being forced to rapidly rotate. The system is very close to Roche lobe overflow and will become a cataclysmic variable in only 70\,Myr.

The opacity from collisionally induced absorption from hydrogen in the ultracool white dwarf atmosphere is strongest in the near-infrared, making this wavelength range particularly sensitive to the temperature and atmospheric composition. Therefore, future measurements of the near-infrared magnitudes for the white dwarf will improve the precision and accuracy of the temperature and composition of the white dwarf.

\section*{Acknowledgments}

We thank the referee, David Pinfield, for his useful comments and suggestions. SGP acknowledges support from the Joint Committee ESO-Government of Chile. ULTRACAM, BTG, TRM, CMC, VSD and SPL are supported by the Science and Technology Facilities Council (STFC). MRS thanks for support from FONDECYT (1100782) and Millennium Science Initiative, Chilean ministry of Economy: Nucleus P10-022-F. The results presented in this paper are based on observations collected at the European Southern Observatory under programme ID 288.D-5015. The Liverpool Telescope is operated on the island of La Palma by Liverpool John Moores University in the Spanish Observatorio del Roque de los Muchachos of the Instituto de Astrofisica de Canarias with financial support from the UK Science and Technology Facilities Council. 

\bibliographystyle{mn_new}
\bibliography{eclipsers}

\label{lastpage}

\end{document}